\def\wgt{\mathop{\rm wgt}}

\def\Hn{\mathcal{H}_2^{\otimes n}}
\def\Pn{\mathscr{P}_{n}}
\def\Prob{\mathop{\rm Prob}}

\def\ket#1{\left|#1\right\rangle}

\documentclass[aps,pra,tightenlines,notitlepage,twocolumn,%
superscriptaddress,
]{revtex4-1}
\usepackage{hyperref}
\usepackage{amsmath}
\usepackage{amsthm}
\usepackage{amssymb}
\usepackage{mathrsfs}
\usepackage{graphicx}
\usepackage{color}
\newtheorem{theorem}{Theorem}
\newtheorem{lemma}[theorem]{Lemma}

\newtheorem{definition}{Definition}
\begin{document}
\title{Thresholds for correcting errors, erasures, and faulty syndrome
  measurements in degenerate quantum codes}

\author{Ilya Dumer}
\affiliation{Department of Electrical Engineering, University of
  California, Riverside, California 92521, USA}

\author{Alexey A. Kovalev}
\affiliation{Department of Physics \&
  Astronomy and Nebraska Center for Materials and Nanoscience,
  University of Nebraska, Lincoln, Nebraska 68588, USA}

\author{Leonid P. Pryadko}
\affiliation{Department of Physics \& Astronomy,
  University of California, Riverside, California 92521, USA}
\date\today

\begin{abstract}
  We suggest a technique for constructing lower (existence) bounds for
  the fault-tolerant threshold to scalable quantum computation
  applicable to degenerate quantum codes with sublinear distance
  scaling.  We give explicit analytic expressions combining
  probabilities of erasures, depolarizing errors, and phenomenological
  syndrome measurement errors for quantum LDPC codes with logarithmic
  or larger distances.  These threshold estimates are parametrically
  better than the existing analytical bound based on percolation.  
\end{abstract}
\maketitle

Quantum computers are (in theory) faster than the classical ones
because of the quantum parallelism.  Instead of working with sets of
classical bits which can store one binary string at a time, a quantum
computer operates with coherent superpositions of exponentially many
basis states encoded in qubits.  Such superpositions are extremely
fragile: decoherence due to environment or intrinsic errors would make
quantum computation unfeasible, were it not for quantum error
correction\cite{shor-error-correct}.  An important result is the
\emph{threshold theorem} stating that with physical qubits and
elementary gates exceeding some accuracy threshold, arbitrarily large
quantum computation is possible with at most polynomial hardware
cost\cite{Shor-FT-1996,Steane-FT-1997,%
  Gottesman-FT-1998,Dennis-Kitaev-Landahl-Preskill-2002,%
  Knill-FT-2003,Knill-2004B,%
  Aliferis-Gottesman-Preskill-2006,Reichardt-2009,%
  Katzgraber-Bombin-MartinDelgado-2009}.

In any quantum error-correcting code (QECC), certain measurements have
to be done repeatedly.  Unlike classical communications setup where
errors happen only during the transmission, this \emph{syndrome}
extraction from a system of qubits is a complicated measurement prone
to errors.  It requires fault-tolerance (FT): all operations have to
be specially designed to limit error propagation.  Requirement of FT
severely restricts the codes which can be used in a quantum computer.
Even though many families of QECCs have been
constructed\cite{Calderbank-1997,Grassl:codetables}, for many years,
FT was demonstrated for only two code families,
concatenated\cite{Shor-FT-1996} and
surface\cite{Dennis-Kitaev-Landahl-Preskill-2002} codes (as well as
related color
codes\cite{Bombin-MartinDelgado-2006,Bombin-2007,Landahl-2011}).  Both
families require substantial hardware overhead, in technical terms,
they have asymptotically zero rates\cite{Bravyi-Poulin-Terhal-2010}.

So far the only family of QECCs where \emph{finite} rates and an FT
threshold to scalable quantum computation are known to coexist are the
quantum LDPC codes\cite{Postol-2001,MacKay-Mitchison-McFadden-2004}.
These are just stabilizer codes\cite{gottesman-thesis,Calderbank-1997}
where stabilizer generators (operators to be measured during QEC)
involve a limited number of qubits each.  Several finite-rate families
of such codes are known\cite{Tillich-Zemor-2009,Kovalev-Pryadko-2012,%
  Andriyanova-Maurice-Tillich-2012,%
  Kovalev-Pryadko-Hyperbicycle-2013,Bravyi-Hastings-2013}.  The
threshold existence has been proved\cite{Kovalev-Pryadko-FT-2013} by
two of us using ideas from percolation theory.  Subsequently, a
related approach has been used by
Gottesman\cite{Gottesman-overhead-2013} who demonstrated that with
such codes, scalable quantum computation is possible with a finite
overhead per logical qubit.

While the technique in Ref.~\onlinecite{Kovalev-Pryadko-FT-2013} shows
the existence of a finite threshold for certain quantum LDPC codes,
the actual threshold value and its dependence on the parameters are
both far off.  The technique\cite{Kovalev-Pryadko-FT-2013} is also too
restrictive: it fails to give a finite threshold whenever
a single qubit is  shared by many stabilizer generators.

In this work we present an approach resulting in a parametrically more
accurate lower bound for the threshold, both in the setting of a
quantum channel and in the FT setting using a phenomenological error
model.  We consider quantum LDPC codes whose distances scale as or
faster than a logarithm of the code length $n$, while all stabilizer
generators are limited to some fixed number $w$ or fewer qubits. For any
sequence of such codes, we
give an analytical lower (existence) bound combining uncorrelated
qubit erasures, depolarizing errors, and syndrome measurement errors.
We also give a similar bound tailored for CSS codes.  These bounds no
longer require that every qubit be included in a limited number of
stabilizer generators.  Tying our lower bound on erasure threshold
with other results\cite{Delfosse-Zemor-2012,Pastawski-Yoshida-2014},
we restrict the parameter space for codes with certain properties.
This approach could also help analyzing FT for other degenerate code
families, e.g., constructed recently by Bravyi and
Hastings\cite{Bravyi-Hastings-2013}.

We consider QECCs
defined on the $n$-qubit Hilbert space $\Hn$, where
$\mathcal{H}_{2}$ is the single-qubit complex Hilbert space spanned by
two orthonormal states $\{\ket0, \ket1\}$.  Any operator acting in
$\Hn$ can be represented as a linear combination of \emph{Pauli
  operators}, elements of the $n$-qubit Pauli group $\Pn$ of size
$2^{2n+2}$,
\begin{equation}
  \Pn=i^{m}\{I,X,Y,Z\}^{\otimes n},\; m=0,\ldots,3,
  \label{eq:PauliGroup}
\end{equation}
where $X$, $Y$, and $Z$ are the usual Pauli matrices, $I$ is the
identity matrix, and $i^{m}$ a phase factor.  \emph{Weight} $\wgt E$
of a Pauli operator $E\in\Pn$ is the number of non-identity terms in
its expansion~(\ref{eq:PauliGroup}).  
A \emph{stabilizer code} $\mathcal{Q}$ with parameters $[[n,k,d]]$ is
a $2^k$-dimensional subspace of the Hilbert space $\Hn$.
$\mathcal{Q}$ is a common $+1$ eigenspace of operators in an Abelian
\emph{sta\-bilizer} group $\mathscr{S}=\left\langle
  G_{1},\ldots,G_{r}\right\rangle $, $-\openone\not\in\mathscr{S}$,
with $r\equiv n-k$ generators $G_i$,%
\begin{equation}
  \label{eq:stabilizer-code}
  \mathcal{Q}=\{\ket\psi: S\ket\psi=\ket\psi,\forall S\in \mathscr{S}\}.
\end{equation}
A more narrow set of Calderbank-Shor-Steane (CSS) codes
\cite{Calderbank-Shor-1996,Steane-1996} contains codes whose
stabilizer generators can be chosen as products of only Pauli $X$ or
Pauli $Z$ operators each.  
For a stabilizer group with $r$ independent generators, the
dimension of the quantum code is given by $k=n-r$; for a CSS code with
$r_X$ generators of $X$-type and $r_Z$ generators of $Z$ type we have
$k=n-r_X-r_Z$.

The error correction is done by measuring the stabilizer generators
$G_i$, $i=1,\ldots, r$; the corresponding eigenvalues $(-1)^{s_i}$,
$s_i\in\{0,1\}$ form the \emph{syndrome} $\mathbf{s}\equiv
(s_1,s_2,\ldots,s_r)$ of the error.  Measuring the syndrome projects
any state $\ket\psi\in\Hn$ into one of the $2^r$ subspaces ${\cal
  Q}_\mathbf{s}$ equivalent to the code ${\cal Q}\equiv {\cal
  Q}_\mathbf{0}$.  An error $E\in\Pn$ is called detectable if it
anticommutes with any generator of the stabilizer; otherwise it is
called undetectable.  Then, for any $\ket\psi\in{\cal Q}$, the
syndrome measured in the state $E\ket\psi$ is non-zero for a detectable
error and it is zero otherwise.  While operators in the stabilizer
group are undetectable, they act trivially on the code; such errors
can be ignored.  Any two
Pauli errors $E_1$, $E_2$ which differ by a phase and an element of
the stabilizer, $E_2=e^{i\alpha}E_1S$, $S\in\mathscr{S}$, are called
degenerate.  Mutually degenerate errors act identically on the code,
they cannot (and need not) be distinguished.

The distance $d$ of the code ${\cal Q}$ is given by the minimum weight
of an undetectable Pauli error $E\in\Pn$ which is not a part of the
stabilizer, $E\not\in \mathscr{S}$.  A code with distance $d$ can
detect any Pauli error of weight up to $d-1$, and it can correct any
Pauli error of weight up to $\left\lfloor d/2\right\rfloor$.  

A code is called \emph{degenerate} if its stabilizer includes a
non-trivial operator $S\in\mathscr{S}$ with weight smaller than the
distance, $0\neq \wgt S<d$.  There is an obvious advantage in choosing
generators of small weight as it simplifies the corresponding quantum
measurements.  Even though with fault-tolerant measurement protocols
one could measure operators involving many qubits (e.g., in the case
of concatenated codes\cite{Shor-FT-1996}), it is much easier to
measure operators which involve only a few qubits.  Thus, we expect
any large quantum code of any use to be degenerate.  The ultimate case
of degeneracy are $w$-limited \emph{quantum LDPC codes}, where every
stabilizer generator involves no more than $w$ qubits.

Existence of a finite error correction threshold requires an infinite
code family with divergent distances.  For example, in codes with a
finite \emph{relative distance} $\delta\equiv d/n$ at large $n$,
uncorrelated single-qubit errors occurring with probabilities
$p<\delta/2$ can be corrected with certainty.  The subject of this
work are codes with \emph{sublinear distance}
scaling\cite{Tillich-Zemor-2009,Kovalev-Pryadko-2012,%
  Kovalev-Pryadko-Hyperbicycle-2013,Andriyanova-Maurice-Tillich-2012},
e.g., power-law $d\propto n^\alpha$, with $\alpha<1$.  Here, at large
$n$, any likely error will have weight $pn$ which is much bigger than
the distance.  

We consider three simple error
models\cite{Bennett-DiVincenzo-Smolin-1997}: quantum depolarizing
channel, where with probability $p$ an incoming qubit is replaced by a
qubit in a random state, without notifying the observer; independent
$X/Z$ errors, where Pauli operators $X$ and $Z$ are applied to each
qubit with probabilities $p_X$ and $p_Z$, respectively, and the
quantum erasure channel, where with probability $y$ each qubit is
replaced by an ``erasure state'' $\ket2$ orthogonal to both $\ket0$
and $\ket1$.  We will also consider FT using phenomenological error
model where measurement errors happen independently with probability
$q$.  Such an error just results in the syndrome bit measured
incorrectly; it does not affect the state of the qubits.

Below we consider infinite sequences of quantum codes whose distances
scale with $n$ at least logarithmically,
\begin{equation}
  \label{eq:log-distance}
  d\ge D\ln n,\quad D>0.
\end{equation}
Super-logarithmic scaling of the distance (including a power law $d\ge
A n^{\alpha}$ with $A,\alpha>0$) gives $D\to\infty$.  We summarize the
constructed thresholds as follows:
\begin{theorem}
  Any sequence of quantum codes (\ref{eq:log-distance}) with
  stabilizer generators of weights $w$ or less can be decoded with a
  vanishing error probability if channel probabilities $(y,p)$ for
  erasures and depolarizing errors satisfy $2(w-1)\,\Upsilon(y,p)\le
  e^{-1/D} $, where%
  \begin{equation}
    \label{eq:threshold-depolarizing}
   \Upsilon(y,p)\equiv y+(1-y)\left\{{2p\over 3}+2\left[{p\over
          3}(1-{p})\right]^{1/2}\right\}.
  \end{equation}%
  \label{th:threshold-stabilizer-w}%
\end{theorem}
\begin{theorem}
  Any sequence of CSS codes (\ref{eq:log-distance}) with generator
  weights not exceeding $w_X$, $w_Z$ can be decoded with  vanishing
  error probabilities 
  if channel probabilities 
  $(y,p_X,p_Z)$ 
  for era\-sures and independent $X/Z$ errors satisfy 
  $(w_X-1)\,\Upsilon_\mathrm{CSS}(y,p_{Z})\le e^{-1/D}$,
  $(w_Z-1)\,\Upsilon_\mathrm{CSS}(y,p_{X})\le e^{-1/D}$,
  where%
  \begin{equation}
    \label{eq:threshold-CSS}
    \Upsilon_\mathrm{CSS}(y,p)\equiv 
    y+2(1-y)\left[{p}(1-p)\right]^{1/2}.
  \end{equation}%
  \label{th:threshold-CSS-w}%
\end{theorem}%
 FT case gives weaker versions of
Theorems \ref{th:threshold-stabilizer-w} and
\ref{th:threshold-CSS-w}: 
\begin{theorem}
  \label{th:threshold-FT}
With the addition of phenomenological syndrome
  measurement errors with probability $q$, vanishing error rates are
  achieved if (\textbf{a}) error probabilities for stabilizer codes in
  Theorem \ref{th:threshold-stabilizer-w} satisfy
  \begin{equation}
    \label{eq:threshold-depolarizing-FT}
  4 \left[q(1-q)\right]^{1/2}+ 2w Y(y,p)\le e^{-1/D},
  \end{equation}
  (\textbf{b}) error probabilities for CSS codes in Theorem
  \ref{th:threshold-CSS-w} satisfy%
  \begin{equation}
    \label{eq:threshold-CSS-FT}
    \begin{aligned}[c]
      4 \left[q(1-q)\right]^{1/2}+ w_X
      Y_\mathrm{CSS}(y,p_{Z})\ge  e^{-1/D},\\
      4 \left[q(1-q)\right]^{1/2}+ w_Z
      Y_\mathrm{CSS}(y,p_{X})\ge  e^{-1/D}.
    \end{aligned}
  \end{equation}  
\end{theorem}

Our analysis is based on counting irreducible undetectable operators:
\begin{definition}
  \label{def:irreducible}
  For a given stabilizer code ${\cal Q}$, an undetectable operator is
  called irreducible if it cannot be decomposed as a product of two
  undetectable Pauli operators with support on non-empty disjoint sets of
  qubits.
\end{definition}
This definition implies:
\begin{lemma}
  \label{th:zero-syndrome-decomposition}
  Any undetectable operator $E\in\Pn$ can be written as $E=\prod_i
  J_i$, where undetectable operators $J_i\in\Pn$, $\wgt J_i\neq0$, are irreducible and pairwise disjoint.
\end{lemma}
For a given code, let $\mathscr{U}\subset\Pn\setminus \mathscr{S}$
denote the set of all non-trivial irreducible undetectable Pauli
operators.

Given some error probability function $P(E)$, consider a syn\-drome-based
decoder which returns the Pauli operator $E\in\Pn$ that produces the
given syndrome and maximizes $P(E)$.  Notice that this is not a true
maximum-likelihood (ML) decoder since we are ignoring contributions of
errors degenerate with $E$.  Using an analogy with statistical
mechanics\cite{Dennis-Kitaev-Landahl-Preskill-2002,Kovalev-Pryadko-SG-2013},
ML decoding corresponds to minimizing the free energy; here we ignore
entropy contribution resulting from degenerate errors and just
minimize the \emph{energy} $\varepsilon(E)\equiv -\ln P(E)$.  Such a
procedure is intrinsically sub-optimal; thus a lower bound for
decoding threshold we get is also a lower bound for syndrome-based ML
decoding.

Now, let $E\in\Pn$ be an error that actually happened, and $E'$ be the
same-syndrome Pauli operator which minimizes the energy
$\varepsilon(E)$.  This error can in principle be found, e.g., by an
exhaustive search.  The product $E' E^\dagger$ is undetectable, it
satisfies Lemma \ref{th:zero-syndrome-decomposition}, which gives a
decomposition $E' E^\dagger=\prod_i J_i$ into irreducible undetectable
operators, $J_i\in\mathscr{S}\cup\mathscr{U}$.  Since the operators
$J_i$ are mutually disjoint, none of them can decrease the energy of
$E'$, $\varepsilon (J_iE')\ge \varepsilon( E')$.  Otherwise $E'$ would
not be the smallest-energy error with the same syndrome.  Thus found
minimal-energy error $E'$ is correct iff $E'E^\dagger $ is trivial,
which implies that all irreducible components must be members of the
stabilizer, $J_j\in\mathscr{S}$ (up to a phase).

Otherwise, there is an irreducible operator $U\in\mathscr{U}$ which
does not increase the energy of the original error $E$, $\varepsilon
(UE) \le \varepsilon(E)$.  Let ${\cal B}(U)\equiv \left\{E\in \Pn:
  \varepsilon (U\!E) \le \varepsilon(E)\right\}$ be the full set of
such ``bad'' errors for a given $U\in\mathscr{U}$.  Minimum-energy
decoding gives vanishing error rate if
\begin{equation}
  \label{eq:min-E-condition}
\Prob\left[E:E\in \bigcup\nolimits_{U\in\mathscr{U}} {\cal
    B}(U)\right]\to 0,\quad n\to\infty.  
\end{equation}
We bound the probability~(\ref{eq:min-E-condition}) by
the sum of probabilities to encounter a ``bad'' 
error from each ${\cal B}(U)$; this gives the
following sufficient condition for error-free decoding:%
\begin{equation}
  \label{eq:succesful-decoding}
  \sum_{U\in\mathscr{U}} \Prob\left[E: E\in {\cal B}(U)\right]\to
  0, \quad n\to\infty.  
\end{equation}
 
For uncorrelated errors only the qubits in the support of $U$ affect
the probabilities in Eq.~(\ref{eq:succesful-decoding}).  Furthermore,
with uniform error distributions, these probabilities depend only on
the weights $m\equiv \wgt U$ of the operators $U\in\mathscr{U}$.  For
example, in the case of erasures with single-qubit probability $y$, a
bad error must cover the entire support of $U$, which gives simply
$\Prob[E:E\in \mathcal{B}(U)]=y^{\wgt(U)}$.  Let $N_m$ denote the number of operators
$U\in\mathscr{U}$ of weight $m\equiv\wgt U$.  Since members of the
stabilizer group are excluded from $\mathscr{U}$, $N_m=0$ for $m<d$.
Thus, in the case of the erasure channel, the condition
(\ref{eq:succesful-decoding}) is equivalent to
\begin{equation}
  \label{eq:succesful-decoding-depolarizing}
  \sum_{m\ge d}N_m y^m\to 0, \quad n\to\infty.
\end{equation}

To construct an upper bound for $N_m$, we use a simplified version of
the cluster-enumeration algorithm originally designed for finding the
distance of a quantum LDPC
code\cite{Kovalev-Dumer-Pryadko-ITA-2013,Dumer-Kovalev-Pryadko-2014}.
Let us assume that the $r$ stabilizer generators $G_i$ are ordered by
weight, $\wgt G_i\le \wgt G_{i+1}$, $1\le i<r$.  Start by placing
either of $ \{X,Y,Z\}$ at a position $j\in \{0,\ldots,n-1\}$ and place
the corresponding Pauli operator as the only element of the list of
the components of the operator being constructed.  At every subsequent
step, take the generator $G_i$ corresponding to a non-zero syndrome
bit with the smallest index $i$, and choose any position $j$ in the
support of $G_i$ that is
not yet selected; there are up to $\wgt G_i-1$
choices.  Choose a single-qubit Pauli different from the term present
at the position $j$ in the expansion (\ref{eq:PauliGroup}) of $G_i$,
and add it to the list.  This sets the syndrome bit $s_i$ to zero
without modifying any of the existing entries in the list.  At every
step of the recursion, zero syndrome means a completed undetectable
cluster; no available positions to correct a chosen syndrome bit means
recursion got stuck.  In either case we need to go back one step by
removing the element last added to the list.  The procedure stops when
we exhaust all choices.

\begin{figure}[htbp]
  \centering
  \includegraphics[width=2.5in]{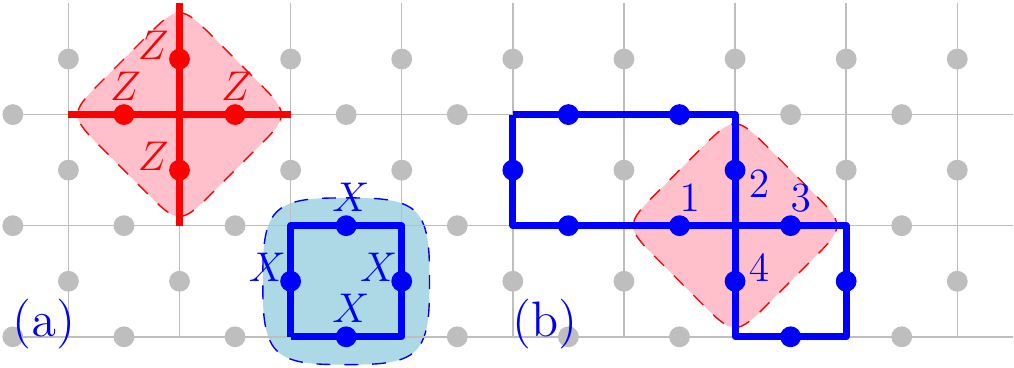}
  \caption{(color online) Structure of the toric code.  (a) Plaquette
    $A_\square$ (shaded rounded square) and vertex $B_+$ (shaded
    diamonds) operators constructed as products of four Pauli $Z$ and
    Pauli $X$ operators respectively.  (b) A reducible cluster (the
    corresponding operator can be split into a product of two
    undetectable operators on non-overlapping subsets) which will be
    counted as one or two clusters depending on the order in which the
    numbered qubits are chosen.}
\label{fig:clust}
\end{figure}

If we limit the recursion to depth $m$, we are only going to construct
operators of weight up to $m$.  There are $3n$ possible choices for
the first step, and up to $2(\wgt G_i-1)$ for each subsequent step.
In the case of a $w$-limited quantum LDPC code, this means no more
than
\begin{equation}
\overline{N}_m=3n [2(w-1)]^{m-1}\label{eq:upper-bound-Nm}
\end{equation}
recursion paths to construct operators of weight up to $m$.  By
construction, the algorithm returns only undetectable operators.
While not all of them are irreducible, it is important that \emph{all
  irreducible operators} of weight $m$ are constructed with depth-$m$
recursion.  Indeed, for a given $U\in\mathscr{U}$, we just have to
start with a non-trivial term in the corresponding
expansion~(\ref{eq:PauliGroup}), and keep choosing only such terms at
every step---the recursion will result in the list corresponding to
$U$ after exactly $m$ steps.  The procedure cannot end earlier since
$U$ is irreducible, and it cannot continue past the $m$-th step since
$U$ is undetectable.  Notice also that we can select
the positions from the support of $U$ in any order as long as $U$
is irreducible.  This is in contrast to the case of an undetectable
but reducible operator, see Fig.~\ref{fig:clust}(b).

These arguments show that $\overline{N}_m$ in
Eq.~(\ref{eq:upper-bound-Nm}) is an upper bound for the number $N_m$
of the irreducible operators $U\in\mathscr{U}$ with weight $\wgt U=m$,
$\overline{N}_m\ge N_m$.

In the case of CSS codes, it is convenient to introduce the sets
$\mathscr{U}_X\subset \mathscr{U}$ and $\mathscr{U}_Z\subset
\mathscr{U}$ of non-trivial irreducible undetectable operators which
are composed only of $X$ and only of $Z$ operators respectively, and
denote $N_m^{(\mu)}$ the number of weight-$m$ operators in
$\mathscr{U}_\mu$, $\mu\in \{X,Z\}$.  For codes in Theorem
\ref{th:threshold-CSS-w}, this gives  improved bounds,
e.g.,%
\begin{equation}
  \label{eq:upper-bound-Nm-CSS}
  N_m^{(X)}\le \overline{N}_m^{(X)}\equiv n(w_Z-1)^{m-1},
\end{equation}
A bound for $N_m^{(Z)}$ can be obtained from
Eq.~(\ref{eq:upper-bound-Nm-CSS}) by exchanging the labels
$X\leftrightarrow Z $.

We illustrate the cluster enumeration procedure on the toric code
$[[2L^2, 2,L]]$, a CSS code with $w_X=w_Z=4$ generators local in two
dimensions.  The qubits are placed on the bonds of an $L\times L$
square lattice with periodic boundary conditions along both bond
directions.  The stabilizer generators are the plaquette and vertex
operators, $A_\square =\prod_{j\in\square} X_j$ and
$B_{\boldsymbol+}=\prod_{j\in{\boldsymbol+}}Z_j$, see
Fig.~\ref{fig:clust}(a).  A type-$X$ cluster can be started by placing
an $X$ operator anywhere, which makes the two operators $B_+$ on the
neighboring vertices unhappy (the corresponding syndrome bits are
non-zero).  Either can be corrected by placing an additional $X$
operator on one of the remaining three open bonds adjoining the
corresponding vertex.  This produces an additional unhappy operator
$B_+$ at the other end of the bond, etc.  An undetectable cluster
corresponds to a closed walk (cycle).  Any cycle can be constructed
this way.  A topologically trivial cycle produces a member of the
stabilizer group, while a cycle winding an odd number of times over
one or both periodicity directions corresponds to a logical operator.
Further, a self-avoiding closed walk corresponds to an irreducible
undetectable operator, while a self-intersecting cycle produces an
operator which can be decomposed into a product of two or more
disjoint cycles, see Fig.~\ref{fig:clust}(b).

Combining Eq.~(\ref{eq:succesful-decoding-depolarizing}) and the bound
$N_m\le \overline{N}_m$, see Eq.~(\ref{eq:upper-bound-Nm}), we can
prove a simplified version of Theorem \ref{th:threshold-stabilizer-w}
for erasure errors only.  Namely, consider the sum
\begin{equation}
  \label{eq:aux-sum}
  Q_d(y)\equiv\sum_{m\ge d}\overline{N}_m y^m={3ny\,[2y(w-1)]^{d-1}\over
    1-2y(w-1)},  
\end{equation}
where we require $2y(w-1)<1$ for absolute convergence.  At large $n$, 
$Q_d(y)$ converges to zero as long as $n [2y(w-1)]^d\to0$.  This is
true for any $y<e^{-1/D}/2(w-1)$ for codes in
Eq.~(\ref{eq:log-distance}).  The
sum~(\ref{eq:succesful-decoding-depolarizing}) is majored term by term
by Eq.~(\ref{eq:aux-sum}).  This proves a version of Theorem
\ref{th:threshold-stabilizer-w} for the erasures only, and gives a
lower bound for erasure threshold, $y_c\ge e^{-1/D}/2(w-1)$.  In the
case of distance scaling as a power-law or faster, the sum
(\ref{eq:aux-sum}) asymptotically vanishes anywhere within the
convergence radius, $y<[2(w-1)]^{-1}$, and we can just set
$e^{-1/D}\to1$.

The proof of Theorem \ref{th:threshold-stabilizer-w} for the combined
erasure and depolarizing errors in the case of generic $w$-limited
quantum LDPC codes, or Theorem \ref{th:threshold-CSS-w} with combined
erasure and independent $X/Z$ errors for CSS codes can be done in a
similar fashion if we notice that the corresponding probabilities in
Eq.~(\ref{eq:succesful-decoding}) can be bounded from above as in
Eq.~(\ref{eq:succesful-decoding-depolarizing}), with some effective
erasure rate $\Upsilon\ge y$.  The complete analysis is given in the 
Appendix \ref{app:erasure-probs}. 

Our arguments so far apply in the conventional ``code-capacity''
setting which assumes that syndrome measurement is done ideally.  In
the case of quantum codes, more important is the fault-tolerant case
where errors may occur at any time during syndrome
measurement\cite{Shor-FT-1996,DiVincenzo-Shor-1996,%
  Knill-error-bound,Gottesman-FT-1998,%
  Steane-1999,Steane-2003,Steane-Ibinson-2005}.  Such a complete
analysis is beyond the scope of this work.  Instead, we give a
simplified estimate based on a phenomenological error model, which
assumes that measured syndrome bits can have errors, but otherwise
there is no effect on the
qubits\cite{Dennis-Kitaev-Landahl-Preskill-2002,Landahl-2011,%
  Kovalev-Pryadko-FT-2013}.  Error correction involves repeated
syndrome measurement cycles and an auxiliary code which combines the
syndromes measured in subsequent cycles.  We only consider the
simplest case where repetition code is used for combining the
syndromes.  For a CSS code, with equal uncorrelated qubit and syndrome
errors $q=p_X=p_Z$, the net effect is equivalent to increasing the
weights of stabilizer generators in Eq.~(\ref{eq:upper-bound-Nm-CSS})
and in Theorem \ref{th:threshold-CSS-w} by two, $w\to w+2$.  With     the surface codes, decoding corresponds to minimal-weight
matching of chains in
three dimensions\cite{Dennis-Kitaev-Landahl-Preskill-2002}.  For a
more general result, we have to bound the number of weight-$m$ clusters
$N_{m,m_q}$ which include $m_q$ ``qubit'' Pauli operators, and $m-m_q$
binary syndrome errors.  Statements of Theorem \ref{th:threshold-FT}
follow from the bound $  N_{m,m_q}\le \overline{N}_{m,m_q}$, 
\begin{equation}
  \label{eq:Nab-bound}
  \overline{N}_{m,m_q}\equiv (3n+r) {m-1\choose
    m_q} w^{m_q} 2^{m-m_q-1}.
\end{equation}
This derivation of this expression and the details of the proof are
given in the Appendix \ref{app:aux-code}.

How tight are the computed bounds?  For the toric code ($w_X=w_Z=4$),
the erasure threshold is $y_c=0.5$ and the ML threshold for
independent $X/Z$ errors is $p_{Zc}=p_{Xc}\approx 0.11$, compared with
$y_c^*=1/3$ and $p_{Zc}^*\approx 0.029$ expected from Theorem
\ref{th:threshold-CSS-w}.  We also checked the accuracy of
Eq.~(\ref{eq:upper-bound-Nm-CSS}) by enumerating irreducible clusters
numerically (see Appendix \ref{app:numerics}) and fitting with $\ln
N_m=A+\zeta_w m$, where $\zeta_w\le w-1$ for CSS codes with row weight
$w$ was expected from Eq.~(\ref{eq:upper-bound-Nm-CSS}).  In
particular, we got $\zeta_6\approx 4.76$, $\zeta_7\approx 5.74$,
$\zeta_8\approx 5.79$ and $\zeta_9\approx 6.78$, indicating that our
 bounds for $N_m$ are relatively tight.

In conclusion, we constructed lower bounds on the thresholds of
weight-limited quantum LDPC codes with sublinear distances scaling
logarithmically or faster with the code length $n$.  These bounds are
based on estimating the number of logical operators which cannot be
decomposed into a product of disjoint undetectable operators.  The
resulting analytical expressions combine probabilities of erasures,
depolarizing errors (independent $X/Z$ errors for CSS codes), and
syndrome measurement errors using a phenomenological error model.
These bounds are much stronger than those constructed
previously\cite{Kovalev-Pryadko-FT-2013}, and they have a different
dependence on the code parameters.  In particular, we no longer
require that each qubit be involved in a limited number of stabilizer
generators.  Qualitatively, the main difference is that the present
analysis is no longer based on percolation theory.

This technique could be applicable not only for weight-limited LDPC
codes, but also for more general degenerate codes, where the
corresponding scaling of $N_m$ can be calculated numerically or
analytically (e.g., in the case of concatenated codes).  It would be
interesting to see if a finite FT threshold exists for finite-rate and
finite-relative distance quantum LDPC codes constructed by Bravyi and
Hastings\cite{Bravyi-Hastings-2013}.  Another potential application
would be the analysis of fault-tolerance of subsystem codes, e.g., a
subclass of those constructed in Ref.\ \cite{Bravyi-subs-2011}.

Our bounds can be also used to limit the parameters of quantum LDPC
codes.  In particular, combining our lower bound $y_c^{\rm (CSS)}\ge
1/(w-1)$ for erasure threshold from Theorem \ref{th:threshold-CSS-w}
with the trivial upper bound $y_c\le (1-R)/2$ suggests that CSS LDPC
codes with super-logarithmic distance scaling do not exist for
$R>1-2/(w-1)$.  In the case of $w=4$ codes this gives $R\le 1/3$,
whereas the only known example of such codes is $R=0$ (toric codes).
These can be further improved by using more accurate upper bounds
constructed specifically for quantum LDPC codes in
Ref.~\onlinecite{Delfosse-Zemor-2012}.

Also, as was pointed to us by Pastawski and Yoshida, our  bounds
on erasure thresholds can be combined with their upper
bound\cite{Pastawski-Yoshida-2014} for codes which include non-trivial
transversal logical gates from $m$-th level of the Clifford
hierarchy\cite{Bravyi-Konig-2013}, $y_m\le 1/m$.  Thus, e.g., only CSS
codes with generators of weight $w\ge m+1$ may include such logical
gates.  Such codes are useful in constructing universal sets of FT
gates acting on logical qubits directly, without the need for
decoding; a set must include at least one non-Clifford operator
($m>2$).  We note that the analysis in Refs.\
\onlinecite{Bravyi-Konig-2013,Pastawski-Yoshida-2014} is largely based
on the cleaning lemma\cite{Bravyi-Terhal-2009,Bravyi-Kitaev-2005}
which utilizes the notion of correctable subsets.  These are
complementary to our irreducible undetectable operators (Def.\
\ref{def:irreducible}); it would be interesting to check if this
relation could help extending the bounds constructed in Ref.\
\onlinecite{Bravyi-Terhal-2009} to general LDPC codes.

\emph{Acknowledgments:} 
This work was supported in part by the
U.S. Army Research Office under Grant No.\ 
W911NF-14-1-0272 (LPP)
and by
the NSF under Grants No.\ PHY-1416578 (LPP), PHY-1415600 (AAK), and 
EPSCoR-1004094 (AAK).
LPP also acknowledges hospitality by
the Institute for Quantum Information and Matter, an NSF Physics
Frontiers Center with support of the Gordon and Betty Moore
Foundation.

\bibliography{lpp,qc_all,more_qc,percol}

\appendix
\begin{widetext}
  \section{Effective erasure probabilities}
  \label{app:erasure-probs}

  Here we derive Eqs.~(\ref{eq:threshold-CSS}) and
  (\ref{eq:threshold-depolarizing}).
  \subsection{CSS code with erasures and independent $X/Z$ errors}

  For a given CSS code, consider an $X$-type undetectable operator
  $U\in\mathscr{U}_X$ of weight $m$.  Fix erasure probability $y$ and
  $X$ qubit error probability $p\equiv p_X$.  We are not concerned
  with $Z$ errors since these do not affect the $Z$-type stabilizer
  generators used to detect the error considered here.  An erasure
  with known location supersedes regular qubit error.  Thus, the
  probability of an error $E$ with $a$ erasures and $b$
  non-overlapping $X$-type errors in the subset of qubits of weight
  $m$ is
  \begin{equation}
    P_E={m\choose a} y^a (1-y)^{m-a} {m-a\choose b} p^b
    (1-p)^{m-a-b}.\label{eq:prob-erasures-xz} 
  \end{equation}
  The probability of an ``inverted'' error $EU$ (notice that this does
  not affect erasures)
$$
P_{EU}={m\choose a} y^a (1-y)^{m-a} {m-a\choose b} (1-p)^b p^{m-a-b}.
$$
The error $E$ is ``bad'', $E\in\mathcal{B}(U)$, if the inverted error
has the same or larger probability; this gives:
\begin{equation}
  (2b+a-m)\ln{1-p\over p}\ge 0.\label{eq:bad-error-CSS}
\end{equation}
Summation of probabilities (\ref{eq:prob-erasures-xz}) gives the net
probability to encounter a ``bad'' error for $U\in \mathscr{U}$, $\wgt
U=m$:
\begin{eqnarray*}
  P_m&=&\sum_{2b+a-m\ge0} P_E(a,b)\\
  &=&\sum_{2b+a-m\ge0} {m\choose a }{m-a\choose b} y^a (1-y)^{m-a}
  [p(1-p)]^{(m-a)/2}\underbrace{\left({p\over
        1-p}\right)^{(2b+a-m)/2} }_{\le 1}
\end{eqnarray*}
The marked term in the last line is smaller or equal than one in the
summation region, the ``bad'' region $\mathcal{B}(U)$.  We make an
upper bound by dropping this term from the product and extending the
summation to all values of $a\ge0$, $b\ge0$ such that $a+b\le m$.  The
summation gives an exponent, with the base of the exponent being the
effective erasure probability in Eq.~(\ref{eq:threshold-CSS}):
\begin{eqnarray*}
  P_m&\le& \sum_{a,b} {m\choose a }{m-a\choose b} y^a (1-y)^{m-a}
  [p(1-p)]^{(m-a)/2}\\
  &=&\sum_{a=0}^m {m\choose a} y^a \left\{2(1-y) [p(1-p)]^{1/2}]\right\}^{m-a}\\
  &=&\left\{y+2(1-y) [p(1-p)]^{1/2}]\right\}^{m}\\ &\equiv&
  [\Upsilon_\mathrm{CSS}(y,p)]^m. 
\end{eqnarray*}

\subsection{Generic stabilizer code with erasures and depolarizing 
  errors}
For a given stabilizer code, consider an undetectable operator
$U\in\mathscr{U}$ of weight $m$. Fix erasure probability $y$ and
depolarizing error probability $p$.  Split the total error weight into
$a$ erasures and $b=b'+b''$ depolarizing errors, with the single-qubit
Pauli errors in $b'$ positions matching those in $U$, and errors in
the remaining $b''$ positions different from the operators in $U$.
Probability of such an error is
\begin{eqnarray}
  \nonumber 
  P_E(a,b',b'')&=&{m\choose a}{m-a\choose b} {b\choose b'}
  y^a(1-y)^{m-a}  
  \left({p\over 3}\right)^{b'}\left(2{p\over
      3}\right)^{b''}(1-p)^{m-a-b'-b''}.\qquad \strut
  \label{eq:prob-erasures-depolarizing}
\end{eqnarray}
The probability of the corresponding ``inverted'' error $EU$ (notice
that the contribution of the $a$ erasures or $b''$ differing positions
remain unaffected):
\begin{eqnarray*}
  P_{EU}(a,b',b'')&=&{m\choose a}{m-a\choose b} {b\choose b'}
  y^a(1-y)^{m-a}  
  \left({p\over 3}\right)^{m-a-b'-b''}\left(2{p\over
      3}\right)^{b''}(1-p)^{b'}. 
\end{eqnarray*}
Respectively, in a bad region $\mathcal{B}(U)$: $P_{EU}\ge P_E$, we
have $2b'+b''+a-m\ge0$.  Using the same trick as before, we have an
upper bound for the total probability of a bad error in a cluster of
size $m$:
\begin{eqnarray*}
  P_m&=& \sum_{E\in\mathcal{B}(U)}P_e \qquad\qquad\left(\text{denote\
    }x\equiv {p/3\over 1-p}\right)\\
  &=& \sum_{E\in\mathcal{B}(U)} {m\choose a}{m-a\choose b} {b\choose b''}
  y^a(1-y)^{m-a}\left(2{p\over
      3}\right)^{b''}  
  x^{(2b'+b''+a-m)/2}\left[\left({p\over
        3}\right)(1-p)\right]^{(m-a-b'')/2}\\
  &\le & \sum_{a,b,b''} {m\choose a}{m-a\choose b} {b\choose b''}
  y^a(1-y)^{m-a}\left(2{p\over
      3}\right)^{b''}  
  \left[\left({p\over
        3}\right)(1-p)\right]^{(m-a-b'')/2}\\
  &=&\sum_{a,b} {m\choose a}{m-a\choose b} 
  y^a(1-y)^{m-a} \left[\left({p\over
        3}\right)(1-p)\right]^{(m-a-b)/2}  
  \left\{2{p\over
      3}+ \left[\left({p\over
          3}\right)(1-p)\right]^{1/2}\right\}^{b} 
  \\ &= &
  \sum_{a,b,b'} {m\choose a}
  y^a(1-y)^{m-a}  
  \left\{2{p\over
      3}+ 2\left[\left({p\over
          3}\right)(1-p)\right]^{1/2}\right\}^{m-a} 
  \\ &=&
  \left(y+(1-y)  \left\{2{p\over
        3}+ 2\left[\left({p\over
            3}\right)(1-p)\right]^{1/2}\right\}\right)^m.
\end{eqnarray*}
The base of the exponent is the effective erasure probability
$\Upsilon(y,p)$, see Eq.~(\ref{eq:threshold-depolarizing}).

\section{Outline the aux code construction for phenomenological error
  model}
\label{app:aux-code}

In this section, we use a binary representation of the Pauli
operators\cite{gottesman-thesis,Calderbank-1997}.  A Pauli operator
can be mapped, up to a phase, to two binary strings,
$\mathbf{v},\mathbf{u}\in\{0,1\}^{n}$,
\begin{equation}
  U\equiv i^{m'}X^{\mathbf{v}}Z^{\mathbf{u}}\:
  \rightarrow(\mathbf{v},\mathbf{u}), 
  \label{eq:mapping}
\end{equation}
where $X^{\mathbf{v}}=X_{1}^{v_{1}}X_{2}^{v_{2}}...X_{n}^{v_{n}}$ and
$Z^{\mathbf{u}}=Z_{1}^{u_{1}}Z_{2}^{u_{2}}...Z_{n}^{u_{n}}$.  A
product of two quantum operators corresponds to a sum $\pmod 2$ of the
corresponding pairs $(\mathbf{v}_i,\mathbf{u}_i)$.  Mapping each
generator $G_j$, $j=1,\ldots,r$ of the stabilizer according to
Eq.~(\ref{eq:mapping}) gives rows of the binary generator matrix
$G=(A_{X}|A_{Z})$, with rows of $A_{X}$ formed by $\mathbf{v}$ and
rows of $A_{Z}$ formed by $\mathbf{u}$ vectors.  For generality, we
also assume that the matrix $G$ may also contain unimportant linearly
dependent rows which are added after the mapping has been done (this
corresponds to adding arbitrary products of stabilizer generators to
the generators of $\mathscr{S}$).  The commutativity of stabilizer
generators corresponds to the following condition on the binary
matrices $A_{X}$ and $A_{Z}$:
\begin{equation}
  A_{X}A_{Z}^{T}+A_{Z}A_{X}^{T}=0 \;\pmod 2.\label{eq:product}
\end{equation}
In the case of the CSS codes, the generator matrix is block-diagonal,
\begin{equation}
  G=\left(\begin{array}{c|c}
      G_{X} & 0\\
      0 & G_{Z}
    \end{array}\right),\label{eq:CSS}
\end{equation}
and the commutativity condition is just $G_{X}G_{Z}^{T}=0$.

It is convenient to introduce the binary check matrix $H\equiv
(A_Z|A_X)$ with the two blocks interchanged.  Then, the commutativity
condition (\ref{eq:product}) becomes simply $HG^T=0$.  Similarly, an
error operator in the form (\ref{eq:mapping}) can be written as a
binary vector $\mathbf{e}^T=(\mathbf{v},\mathbf{u})$; the
corresponding syndrome is just $\mathbf{s}=H\mathbf{e}$.

In the case of repeated syndrome measurements, in the phenomenological
model, qubit errors accumulate while syndrome errors do not.  Thus, if
we denote qubit errors between $(i-1)$\,th and $i$\,th measurement
rounds as $\mathbf{e}_i$, and syndrome errors in these rounds as
${\boldsymbol\epsilon}_i$, $i=1,\ldots,m$, we have the equations
\begin{eqnarray*}
  H\mathbf{e}_1&=&\mathbf{s}_1+{\boldsymbol\epsilon}_1,\\
  H(\mathbf{e}_1+\mathbf{e}_2)&=&\mathbf{s}_2+{\boldsymbol\epsilon}_2,\\
  \ldots \\
  H(\mathbf{e}_1+\mathbf{e}_2+\ldots+\mathbf{e}_m)&=&\mathbf{s}_m+{\boldsymbol\epsilon}_m.  
\end{eqnarray*}
Adding pairs of neighboring equations, and moving syndrome errors to
the left, we obtain the following equations
\begin{eqnarray*}
  H \mathbf{e}_1+{\boldsymbol\epsilon}_1&=&\mathbf{s}_1,\\
  H \mathbf{e}_2+{\boldsymbol\epsilon}_1+{\boldsymbol\epsilon}_2&=&\mathbf{s}_1+\mathbf{s}_2,\\
  \ldots \\
  H \mathbf{e}_{m}+{\boldsymbol\epsilon}_{m-1}+{\boldsymbol\epsilon}_m&=&\mathbf{s}_{m-1}+\mathbf{s}_{m}.
\end{eqnarray*}
We notice that thus constructed combined code is not particularly good
if treated as a classical binary code.  Indeed, a low-weight error
which consists of a single qubit error $\mathbf{e}_s$, $\wgt
\mathbf{e}_s=1$, the syndrome error
${\boldsymbol\epsilon}_s=H\mathbf{e}_s$, and (in the case $s<m$) an
identical single-qubit error $\mathbf{e}_{s+1}=\mathbf{e}_s$, will not
be detected.  On the other hand, for $s<m$, such an error obviously
produces no mistake; we do not need to correct these errors just as we
do not need to correct trivial degeneracy errors.  In the case of
$s=m$, the error cannot be detected at this cycle of measurements; it
is convenient to reassign all such errors to the subsequent cycle
(unless it corrects itself, such an error will be detected in the next
round of measurements).  This prescription is equivalent to setting
$\boldsymbol\epsilon_m=\bf 0$.
 
If we combine the (shortened) error and the syndrome vectors into
unified vectors $\mathbf{e}$ and $\mathbf{s}$, respectively, we can
write these equations simply as $P\mathbf{e}=\mathbf{s}$, where
\begin{equation}
  \label{eq:combined-H}
  P=\left(I_m\otimes H_{r\times n},R_{m\times (m-1)}\otimes
    I_r\right) 
\end{equation}
is a matrix formed by two blocks, with ``$\otimes$ '' denoting
Kronecker product, $I_m$ an $m\times m$ identity matrix, and
\begin{equation}
  \label{eq:R-matrix}
  \quad R_{m\times (m-1)}\equiv \left(
    \begin{array}[c]{ccccc}
      1& \\
      1&1&\\
      & \ddots&\ddots\\
      & & 1&1\\
      & & & 1
    \end{array}
  \right)
\end{equation}
is a transposed check matrix for the length-$m$ repetition code.
Further, it is easy to check that any combination of trivial
degeneracy errors and undetectable self-corrected errors can be
expressed as a linear combination of the rows of the matrix
\begin{equation}
  \label{eq:combined-P}
  Q=\left(
    \begin{array}[c]{cc}
      [R^T]_{(m-1)\times m}\otimes I_n&I_{m-1}\otimes [H^T]_{n\times r}\\
      I_{m}\otimes G_{r'\times n},&0
    \end{array}
  \right).
\end{equation}
Obviously, $PQ^T=0\bmod2$. In fact, the matrices $P$ and $Q$ can be
viewed as CSS generators of a code similar to a hypergraph-product
code\cite{Tillich-Zemor-2009,Kovalev-Pryadko-2012}, with the
difference that one of the constituent codes is a quantum, not a
classical code.  The parameters of such a code can be easily expressed
in terms of those of the length-$m$ repetition code and the original
quantum code with the check matrix $H$ and the generator matrix $G$.
Namely, this code of length $N=mn+(m-1)r$ encodes exactly $K=k$ qubits
with the distance $D=\min(d,m)$.

Such an auxiliary code results in a generalization of the
three-dimensional line-matching for the case of the surface
codes\cite{Dennis-Kitaev-Landahl-Preskill-2002}; a similar
construction has also been discussed in Refs.\
\onlinecite{Landahl-2011,Kovalev-Pryadko-FT-2013}.  For our purposes,
it is important that for an original $w$-limited LDPC code, the check
matrix $P$ has row weight limited by $w+2$, with up to $w$ positions
in the block corresponding to qubit errors, and the remaining one or
two positions in the block corresponding to syndrome measurement
errors.

\section{Effective erasure probabilities with syndrome errors}
\label{app:FT-clusters}

Here, we derive Eqs.~(\ref{eq:threshold-depolarizing-FT}) and
(\ref{eq:threshold-CSS-FT}).  The derivation is similar to that in
Appendix \ref{app:erasure-probs}, with the difference that qubit
errors and syndrome measurement errors have to be treated differently.
Therefore, we consider an error $E$ of the total weight $m$, with
$m_q$ positions in the ``qubit'' part corresponding to the first block
of Eq.~(\ref{eq:combined-H}), and the remaining $m-m_q$ positions in
the ``syndrome'' part.  To simplify the derivation, we omit the
erasures.
\subsection{CSS code with independent $X/Z$ errors: FT case}

Consider an $X$-type binary error $e$ which produces zero syndrome
with the check matrix (\ref{eq:combined-H}) where we only include the
generator $G_Z$, see Eq.~(\ref{eq:CSS}).  The error probabilities for
qubits and syndrome bits are $p\equiv p_X$ and $q$, respectively.
Then, the probability for $e$ to cover $b$ out of $m_q$ qubit
positions and $f$ out of $m-m_q$ syndrome positions is
$$
P_e={m_q\choose b}{m-m_q\choose f}p^{b}(1-p)^{m_q-b}
q^{f}(1-q)^{m-m_q-f},
$$
whereas probability of the same error plus the codeword (error
inverted) is
$$
P_{e+c}= {m_q\choose b}{m-m_q\choose f}(1-p)^{b}p^{m_q-b}
(1-q)^{f}q^{m-m_q-f}.
$$
For a ``bad'' error, the ratio $P_{e+c}/P_e\ge1$, which defines the
bad-error region $\mathcal{B}$:
$$
(2b-m_q)\ln {1-p\over p}+(2f+m_q-m)\ln {1-q\over q}\ge 0.
$$
In the absence of syndrome measurement errors this goes over to $2
b\ge m_q$, cf.\ Eq.\ (\ref{eq:bad-error-CSS}) with $a\to0$.  Now, we
need to find the total probability of an error in $\mathcal{B}$:
\begin{eqnarray*}
  P_\mathrm{bad}&=&\sum_{(b,f)\in \mathcal{B}} P_e(b,f)\\&=&
  \sum_{(b,f)\in\mathcal{B}}
  {m_q\choose b}{m-m_q\choose f}p^{b}(1-p)^{m_q-b}
  q^{f}(1-q)^{m-m_q-f}\\
  &=&
  \sum_{(b,f)\in\mathcal{B}}
  {m_q\choose b}{m-m_q\choose f}\underbrace{\left({p\over 1-p}\right)^{(2b-m_q)/2}
\left({q\over 1-q}\right)^{(2f+m_q-m)/2}}_{\le 1} 
[p(1-p)]^{m_q/2}
[q(1-q)]^{(m-m_q)/2}.
\end{eqnarray*}
The marked portion of the expression does not exceed one in
$\mathcal{B}$; dropping it and extending the summation to all $b\le
m_q$, $f\le m-m_q$, we obtain 
\begin{equation}
  \label{eq:prob-bad-CSS-ab}
  P_\mathrm{bad}(m,m_q)\le 2^{m}[p(1-p)]^{m_q/2} [q(1-q)]^{(m-m_q)/2}.
\end{equation}

We consider specifically the generators coming from the matrix in the
form (\ref{eq:combined-H}), with weight up to $w\equiv w_Z$ in the
qubit positions, and weight up to $2$ in the syndrome positions.
Then, the number of clusters can be bounded by 
$$
N_{m,m_q}\le \overline{N}_{m,m_q}^{(\mathrm{CSS})}=n{m-1\choose
  m_q}w^{m_q}2^{m-m_q-1}\le 
n{m\choose m_q}w^{m_q}2^{m-m_q}, 
$$
cf.\ Eq.~(\ref{eq:Nab-bound}).  This is slightly worse than the bound
(\ref{eq:upper-bound-Nm-CSS}) since we do not know whether the
particular non-zero check bit originated from a qubit or a syndrome
error.  Overall, the net probability for ``bad'' error of combined weight
exceeding the distance $d$ can be bounded by
\begin{eqnarray}
  P_\mathrm{bad}^\mathrm{(tot)}&\le& \sum_{m\ge d} \overline{N}_{m,m_q}^{(\mathrm{CSS})} 
  P_\mathrm{bad}(m,m_q)\nonumber\\
  &\le &n\sum_{m\ge d}{m\choose m_q}2^m
  [w^2p(1-p)]^{m_q/2}[4q(1-q)]^{(m-m_q)/2}\nonumber \\
  &=&n\sum_{m\ge d}\left(4[q(1-q)]^{1/2}+2w[p(1-p)]^{1/2}\right)^m.
  \label{eq:prob-bad-CSS-tot}
\end{eqnarray}
Analysis of the convergence for codes (\ref{eq:log-distance}) gives
the sufficient condition
\begin{eqnarray}
  \label{eq:threshold-curve-CSS}
4[q(1-q)]^{1/2}+  2w[p(1-p)]^{1/2}\le e^{-1/D}, 
\end{eqnarray}
a special case of Eqs.~(\ref{eq:threshold-CSS-FT}) for $y\to0$. 

The derivation of the complete
expressions~(\ref{eq:threshold-CSS-FT}), and of
Eq.~(\ref{eq:threshold-depolarizing-FT}) for the case of combined
erasures, depolarizing errors, and syndrome errors is similar. 
\section{Numerics}
\label{app:numerics}

We implemented the cluster-enumeration algorithm presented in the main
text numerically on Mathematica.  Fig.~\ref{fig:clusterscaling} shows
$N_m$ computed for the hyperbicycle
code\cite{Kovalev-Pryadko-Hyperbicycle-2013} $[[168,6,12]]$, a CSS
code with generators of weight $w_X=w_Z=6$ produced from the cyclic
binary code $[7,3,4]$; a hypergraph-product
code\cite{Tillich-Zemor-2009} $[[1508,100,6]]$ with row weights
limited by $w_X=w_Z=7$ produced from a Gallager code $[32,10,6]$ with
row weights $4$; and two matching three-dimensional codes
$[[2940,6,12]]$ and $[[12568,100,6]]$ with the binary check matrix
(\ref{eq:combined-H}).  The latter codes used $m=12$ and $m=6$ layers
respectively, and have the respective row weights bounded by $w=8$ and
$w=9$.
\begin{figure}[htbp]
  \centering
  \includegraphics[width=3in]{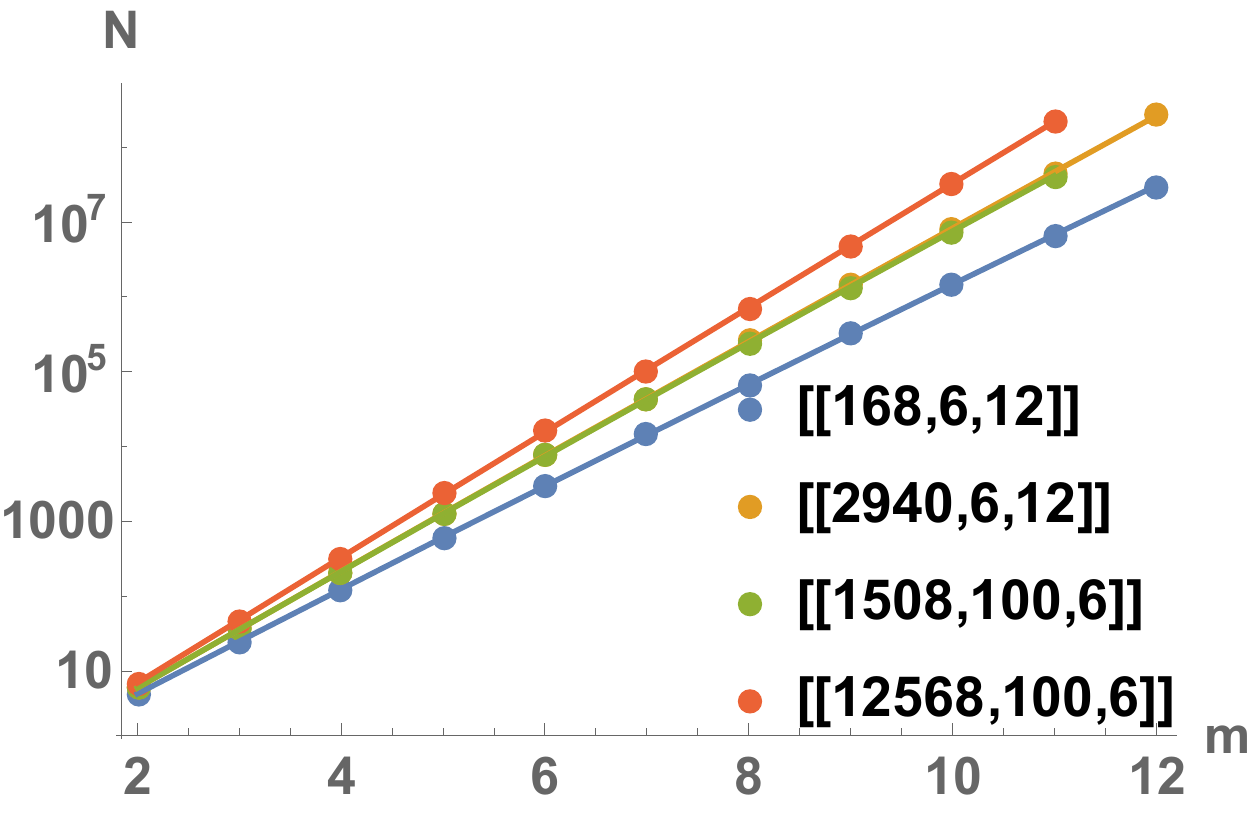}
  \caption{(Color online) Numbers of undetectable clusters computed
    numerically for several codes as indicated.  See text for the
    details of the codes.  The fits $\ln N=A+\zeta m$ give slopes
    $\zeta=4.76261, 5.7921, 5.74431, 6.7889$, respectively. }
  \label{fig:clusterscaling}
\end{figure}
\end{widetext}

\end{document}